\documentclass[aps, prx, twocolumn, article, superscriptaddress, longbibliography]{revtex4-2}

\usepackage{amsmath,amssymb,amsthm}
\usepackage{quantum}
\usepackage{color}
\usepackage{tikz}
\usepackage[utf8]{inputenc}
\usepackage{hyperref}

\begin{document}
\title{Optimal adaptation of surface-code decoders to local noise}

\author{Andrew S. Darmawan}
\affiliation{Yukawa Institute for Theoretical Physics (YITP), Kyoto University, Kitashirakawa Oiwakecho, Sakyo-ku, Kyoto 606-8502, Japan}

\begin{abstract}
    Information obtained from noise characterization of a quantum device can be used in classical decoding algorithms to improve the performance of quantum error-correcting codes. Focusing on the surface code under local (i.e. single-qubit) noise, we present a simple method to determine the maximum extent to which adapting a surface-code decoder to a noise feature can lead to a performance improvement. Our method is based on a tensor-network decoding algorithm, which uses the syndrome information as well as a process matrix description of the noise to compute a near-optimal correction.  By selectively mischaracterizing the noise model input to the decoder and measuring the resulting loss in fidelity of the logical qubit, we can 
determine the relative importance of individual noise parameters for decoding. We apply this method to several physically relevant uncorrelated noise models with features such as coherence, spatial inhomogeneity and bias. While noise generally requires many parameters to describe completely, we find that to achieve near optimal decoding it appears only necessary adapt the decoder to a small number of critical parameters.
\end{abstract}

\date{\today}

\maketitle

\section{Introduction}
The surface code is a promising platform for scalable fault-tolerant quantum computing \cite{dennis_topological_2002, bravyi_quantum_1998, fowler_high-threshold_2009, fowler_surface_2012}.
It has a high threshold compared to other quantum error-correcting codes, and only requires nearest-neighbor interactions on a two-dimensional square lattice to implement. Recent experimental progress suggests that practical surface-code error correction may be realized in the near future \cite{google_quantum_ai_exponential_2021, zhao_realization_2022, acharya_suppressing_2023,  bluvstein_logical_2023}.  

An essential part of surface-code error correction is the decoder. The decoder for the surface code is a classical algorithm that takes the measured syndrome as input and outputs a correction operation to restore the encoded data. 
A variety of surface-code decoding algorithms have been developed, each with strengths and weaknesses \cite{ harrington_analysis_2004, duclos-cianci_fast_2010, wootton_high_2012, bravyi_quantum_2013, duclos-cianci_fault-tolerant_2014, hutter_efficient_2014, watson_fast_2015,herold_cellular-automaton_2015, herold_cellular_2017,  delfosse_almost-linear_2017, tuckett_ultrahigh_2018, sabo_trellis_2022, chubb_statistical_2021, chubb_general_2021}. 
Although the decoder is a purely classical algorithm, improvements in the decoder can drastically reduce the logical error rate, and thereby reduce the amount of physical overhead required for fault tolerance.

A decoder that uses information about the physical noise can correct errors with a substantially higher success rate than one that does not.
Some notable examples of noise-adapted decoders include those adapted to noise bias~\cite{tuckett_ultrahigh_2018, tuckett_tailoring_2019, tuckett_fault-tolerant_2020, bonilla_ataides_xzzx_2021}, correlations between $X$ and $Z$ errors ~\cite{delfosse_decoding_2014, fowler_optimal_2013}, spatial correlations ~\cite{nickerson_analysing_2019, darmawan_linear-time_2018, chubb_statistical_2021, chen_calibrated_2021} and error probabilities on individual circuit elements~\cite{fowler_topological_2012, huang_fault-tolerant_2020}. In addition, information from lower level codes concatenated with surface codes (e.g. concatenated bosonic-qubit surface codes~\cite{fukui_high-threshold_2018, vuillot_quantum_2019}) and `soft' information from measurements \cite{pattison_improved_2021} can also be used by surface-code decoders to improve performance. 

There are, however, practical limitations to noise-adapted decoding. For one, only so much can be known about physical noise. Practical noise characterization methods will typically make assumptions about the noise, and cannot efficiently provide a complete description of the noise \cite{eisert_quantum_2020}. Furthermore, it is not obvious, in general, how to design a sufficiently fast decoding algorithm that incorporates a given noise feature, and whether there will actually be a significant benefit in doing so. Algorithmic considerations are crucial in architectures where run-time must be extremely fast to keep up with the stream of data produced by repeated syndrome measurements \cite{fowler_minimum_2013, das_scalable_2020, holmes_nisq_2020, ueno_qecool_2021-1, das_lilliput_2021}. Hence, understanding the trade-offs between accuracy and runtime of noise-adapted decoding is important for the development of practical quantum error correction \cite{delfosse_how_2023}. 

In this paper, we present a simple method to determine the extent to which information about the noise can improve surface-code decoding performance. 
This allows us to identify which noise parameters are important for decoding, as well as those that can be ignored with little negative impact. 
The performance gain from using noise information that we calculate is optimal over all surface-code decoders, and is therefore a property of the noise and the code, rather than a peculiarity of any particular decoding algorithm (which will typically be suboptimal even with perfect knowledge of the noise). 

We demonstrate this approach on a variety of local noise models. We find that, generally, only small amount of information about the noise is necessary for realizing near optimal decoding. Therefore, a potential practical application of this method is to guide the development of fast decoders and efficient noise characterization methods that are targeted at the noise parameters that most strongly affect performance.

The method is simple and is based on a decoding algorithm previously studied in Ref. \cite{darmawan_linear-time_2018} that can be implemented using tensor-network methods. The decoding algorithm takes as input both the syndrome, and a full description of the physical noise (given by the process matrix) and returns the optimal or near-optimal correction operation for that syndrome consistent with the noise. This is in contrast to more commonly studied tensor-network decoders that typically take only Pauli error probabilities as input \cite{bravyi_efficient_2014-1} and don't include coherent parts of the noise. We call the noise model input to the decoder the \emph{decoder noise model}. Since the decoder effectively implements a kind of density matrix time evolution algorithm, we will henforth refer to it as a \emph{simulation-based decoder}, and we review how this works in Sec. \ref{s:tndecoder}. 

This decoder will achieve near-optimal performance when the decoder noise model is exactly the same as the physical noise. However, we also consider the case where the decoder is miscalibrated. The simulation-based decoder is miscalibrated if the decoder noise model differs from the physical noise. Such a mismatch represents the situation where the decoder is not perfectly adapted to the noise, or is provided with a limited, or imperfect description of the physical noise. In general, the performance of the decoder will decrease as the difference between the decoder noise model and the physical noise model increases.

As an example of decoder miscalibration, we could have physical noise that is inhomogeneous (i.e. varies across qubits) but the decoder takes a homogeneous noise model as input. Another example could be where the physical noise is non-Pauli noise (e.g. a coherent unitary rotation), but the decoder takes a Pauli noise description as input. 

If the performance of the miscalibrated decoder is practically indistinguishable from the optimal decoder, it shows that the decoder does not need to be adapted to the mischaracterized noise feature in order to achieve optimal performance. For instance, if the physical noise is a unitary rotation, yet optimal performance is achieved when the decoder noise model is the Pauli twirl of the physical noise, it shows that the decoder only needs to be adapted to the incoherent component of the noise.
Conversely, if a significant difference in performance is observed between the optimal and miscalibrated decoders, then we can conclude that the mischaracterized noise parameter is important to decoding. 

Any parameter in the process matrix description of the noise translates directly to a parameter in the simulation-based decoder, allowing us to observe the effect of adapting the decoder to specific noise parameters. Note that for heuristic decoding algorithms, like minimum-weight perfect matching (MWPM), the performance is usually suboptimal even when the decoder is perfectly calibrated to the noise. For instance, even if edge weights in MWPM are set using the exact same parameters as the physical noise, the performance will in general be lower than an optimal decoder\cite{lange_data-driven_2023}. One reason for this is that MWPM cannot take into account correlations between X and Z errors. Therefore, for MWPM or other heuristic algorithms, it is not possible to separate the effect of noise adaptation from particular shortcomings of the decoding algorithm. This justifies the use of simulation-based decoding for this purpose.

We have used Pauli simulations, tensor-network simulations and state vector simulations depending on the type of noise under consideration. The simulation-based decoding algorithms used for different noise models are described in Secs. \ref{s:tndecoder} and \ref{s:optdecoder}. We have restricted to relatively small system sizes: $d=9$ for the tensor-network simulations and $d=3$ for the state vector simulations. On these sizes optimal decoding can be implemented exactly. While it is possible to scale the decoders and simulators to larger sizes by introducing approximations ~\cite{darmawan_tensor-network_2017, darmawan_linear-time_2018}, this can introduce errors, and was unnecessary to demonstrate our method.

\subsection{Summary of main results}

In Sec. \ref{s:inhomogeneity} we study the extent to which adapting a decoder to spatial inhomogeneity can improve performance. We investigate a situation that is typical for superconducting qubits, where energy relaxation and phase relaxation times ($T_1$ and $T_2$ respectively) vary across different qubits in the device. Our numerical simulations show that large performance improvements are possible with a decoder that takes this inhomogeneity into account, compared to a decoder that only knows the average $T_1$ and $T_2$. We can also obtain a more refined understanding of what noise parameters need to be known for optimal decoding. By mischaracterizing the overall noise strength on each qubit, while keeping the ratios $T_1/T_2$ correct, the performance is again found to be near optimal, suggesting that it is inhomogeneity in the nature of the noise (phase damping or amplitude damping) that matters to the decoder, rather than variation in noise strength across qubits in the device. 

In Sec \ref{s:coherence} we turn our attention to coherent noise. We investigate whether using a full characterization of a coherent noise model in the decoder has an advantage over only knowing the incoherent components (the Pauli error probabilities). While some improvement is observed when adapting the decoder to coherence, our results suggest that in general, the benefit of is small and the effect coherence is better taken care of with other methods, rather than decoding. Since the majority of noise parameters in a CPTP map describe `coherence' (i.e. off diagonal terms in the $\chi$ matrix), this result suggests that optimal decoding may be possible with a highly compressed description of the noise.

In Sec. \ref{s:bias} we investigate the importance of bias in surface-code decoding. It has previously been shown that surface code performance can be substantially improved by adapting the decoder to noise bias \cite{tuckett_ultrahigh_2018, tuckett_tailoring_2019, tuckett_fault-tolerant_2020, bonilla_ataides_xzzx_2021}. Here we provide insight into the kind of noise information required by the decoder to achieve optimal performance. We demonstrate that surface-code decoder performance is more sensitive to miscalibration for biased noise models than unbiased noise. For biased noise both the amount of bias, as well as the strength of the noise, must be known to achieve optimal performance, however the characterization does not have to be very accurate.

Finally, in Sec. \ref{s:generic} we present a more general method for determining the critical noise parameters for decoding, which can be applied to any local noise model. The method involves perturbing the decoder noise model in various directions away from the optimal, and comparing the loss in fidelity observed in each case. We demonstrate the method on homogeneous amplitude-phase damping noise, and observe that the critical noise parameter corresponds in this case to the correlations between the $X$ and $Z$ Pauli errors.

\section{Method}
\label{s:method}

Here we review the surface code and explain optimal, or simulation-based, decoding. We also describe a how to use optimal decoding to identify noise parameters that are critical to decoding.  The single-qubit Pauli operators are denoted by $X,Y,Z$ and we use a subscript to label the qubit on which an operator acts non-trivially e.g. in $X_i$  is the operator that acts as $X$ on qubit $i$ and as the identity elsewhere.  

\subsection{Surface code}

\begin{figure}
    \includegraphics[width=0.35\textwidth]{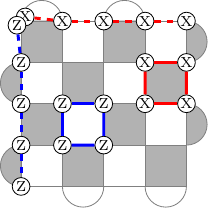}
    \caption{Surface code layout where qubits are located on the vertices of a square lattice and check operators correspond to faces. Check operators are products of Pauli $X$ around grey faces and Pauli $Z$ around white faces. Logical $Z$ is a string of Pauli $Z$ connecting top to bottom, and logical $X$ is a product of Pauli $X$ from connecting left to right, as illustrated.}
    \label{f:layout}
\end{figure}

The surface code is defined with respect to a set of commuting check operators, which act locally on qubits arranged on the vertices of a square lattice. We have illustrated the check operators in Fig. \ref{f:layout}. The $X$ check operators are defined as $A_v = \bigotimes_{i\in v} X_i$, where the $v$ subscript runs over all grey faces and the product runs over qubits on the vertices of $v$. Analogously, the $Z$ check operators are given by $B_p = \bigotimes_{i\in p } Z_i$ where $p$ runs over white faces. The code space is defined as the simultaneous ${+}1$ eigenspace of all checks. The group generated by all such checks is called the stabilizer. In this work we consider the surface code with open boundary conditions described in Ref. \onlinecite{bombin_optimal_2007}, which has two-qubit checks on the boundary. 

 With these boundary conditions, the code space is 2-dimensional, i.e. it encodes a single qubit. The logical $Z$ operator $\overline{Z}=\bigotimes_{i} Z_i$ consists of a string of $Z$ operators connecting the top left and bottom left boundaries of the lattice, while the logical $X$ operator $\overline{X}=\bigotimes_{i } X_i$ consists a string of $X$ operators connecting the top left and top right boundaries of the lattice, as illustrated in Fig. \ref{f:layout}. Any deformation of these strings obtained buy multiplying them by a stabilizer element has equivalent action of the code space. The logical operators commute with every check, and so preserve the code space, yet they act non-trivially on states encoded within the code space. We define the logical qubit states $\ket{0}_L$ and $\ket{1}_L$ as the ${+}1$ and ${-}1$ eigenstates of $\overline{Z}$ within the code space, which implies that $\overline{Z}$ and $\overline{X}$ act as Pauli $Z$ and $X$ operators on the logical qubit. The code distance $d$ is defined as the lowest weight logical operator, which on a square lattice is $\sqrt{N}$ where $N$ is the number of qubits in the code.

\subsection{Optimal surface-code decoding}
\label{s:decoding}

In this section we will outline the main idea of simulation-based decoding. In particular, we explain how calculating the logical channel is sufficient for optimal decoding, following Ref. \cite{darmawan_linear-time_2018}. Computing the logical channel is, in general, challenging, however, we explain in the following section how it can be computed for the surface code using tensor-network methods. 
We remark that certain conventions about, for instance, the normalization of the logical channel and the definition of the logical error rate are different from Ref. \cite{darmawan_linear-time_2018}, although the essential ideas remain the same.

Let $\mathcal{N}$ be an $N$-qubit noise map, which we assume is completely positive and trace preserving (CPTP). In the case of spatially uncorrelated noise we can write $\mathcal{N} = \bigotimes_{i=1}^N \mathcal{N}_i$ where $\mathcal{N}_i$ is a CPTP map that acts non-trivially on qubit $i$ and as the identity elsewhere. Assume, for now, that check measurements are performed noiselessly, and the projection onto the syndrome subspace corresponding to syndrome $s$ is given by $\Pi_s$. 

For a given syndrome $s$ of the rotated surface code there are four inequivalent Pauli recovery operators that return the post-measurement (unnormalized) state $\Pi_s\circ \mathcal{N} (\rho_{\rm in})$ to the code space. These recovery operators differ on the logical space by multiplication by a logical Pauli operator $L \in \{\overline{I}, \overline{X}, \overline{Y}, \overline{Z}\}$. One such correction, which we call the fiducial correction $f_s$ is the Pauli operator that connects every flipped $Z$ check to the left boundary by a string of Pauli $X$ operators, and each flipped $X$ check to the top boundary by a string of Pauli $Z$ operators. We define the recovery operator $\mathcal{R}_s := f_s \circ \Pi_s$, which effectively maps the noisy state back to the code space. 

The map $\mathcal{R}_s\circ \mathcal{N}$ for the surface code is a single qubit map from the code space to itself and can be expressed using a $4\times4$ process matrix or  $\chi$ matrix (which are defined in Sec. \ref{s:coherence}). Note that $f_s$ is simply an operator that returns the post-measurement state to the code space, and it is not chosen to minimise the probability of a logical error. Minimising the logical error rate is done by the decoder correction $\mathcal{D}(s) \in \{\overline{I}, \overline{X}, \overline{Y}, \overline{Z}\}$ which can be thought of as being applied after the recovery (although in practice only needs to be stored in classical memory, to keep track of the Pauli frame of the encoded qubit). 

In the case of ideal measurements, we thus describe the evolution of the encoded state through a round of error correction with the completely positive trace non-increasing map
\begin{equation}
    \mathcal{E}_s(\rho) := \mathcal{D}(s) \circ \mathcal{R}_s \circ \mathcal{N}(\rho)\,,
\end{equation}
which we call the logical channel. 
The decoder correction $\mathcal{D}(s)$ is selected by a decoding algorithm based on the input syndrome and information about the noise. 
Given that we want to minimise the disturbance on the encoded state, the optimal decoder correction, which we denote by $\mathcal{D}_{\rm opt}$, is the one that minimises the distance between $\mathcal{E}_s$ (normalized) and the identity, i.e. 
\begin{equation}
    \mathcal{D}_{\rm opt}(s):=\underset{L\in \{\overline{I}, \overline{X}, \overline{Y}, \overline{Z}\}}{\rm argmin}||\fr{p_s} L \circ \mathcal{R}_s \circ \mathcal{N}-I||_\diamond\,,
    \label{e:optimal}
\end{equation}
where $||.||_\diamond$ is the diamond norm, and the normalization factor $p_s:={\rm Tr}(\Pi_s \mathcal{N}(\fr{2}I))$ is the probability of measuring the syndrome $s$ given the noise model $\mathcal{N}$ and maximally mixed\footnote{Note that in general, the probability of obtaining syndrome $s$ depends on the encoded state, however for the sake of computing Eq. \eqref{e:optimal} using the operator distance we always use the normalization $p_s:={\rm Tr}(\Pi_s \mathcal{N}(\fr{2}I))$ (i.e. we assume a completely unknown encoded state)} encoded state $\fr{2}I$. Note that noisy syndrome extraction may also be included in this picture simply by replacing $\mathcal{R}_s\circ \mathcal{N}$ with other dynamics (such as a noisy syndrome extraction circuit), as we illustrate in Sec. \ref{s:optdecoder}. 

 In Refs. \onlinecite{darmawan_tensor-network_2017,darmawan_linear-time_2018} we showed that, in the case of noiseless syndrome extraction, a tensor-network method could be used to compute the $4\times 4$ process matrix of the single-qubit logical channel $\mathcal{R}_s\circ\mathcal{N}$ for any syndrome $s$ and local noise map $\mathcal{N}$. Then, by computing the diamond norm distance in  Eq. \eqref{e:optimal} for all four decoder corrections $L\in \{\overline{I}, \overline{X}, \overline{Y}, \overline{Z}\}$, the optimal correction $\mathcal{D}_{\rm opt}$ can be chosen simply as the minimal one. Note that, for Pauli noise, the $\chi$ matrix description of the logical channel $\mathcal{R}_s\circ \mathcal{N}$ is diagonal and the diagonal entries are proportional to the coset probabilities calculated using the maximum likelihood decoder of Ref. \cite{bravyi_efficient_2014-1}.

The tensor-network algorithm that we use to calculate the logical channel (described in more detail in Sec. \ref{s:tndecoder}),  has runtime linear in the system size if the bond dimension used in the approximate tensor-network contraction is held constant. While the exact implementation of this algorithm has exponential runtime scaling, it can nevertheless reach reasonably large system sizes (of well over 100 qubits), which is sufficient to show size dependence in many cases.

Given a noise map and a decoder, we define the logical error rate as the average diamond distance of the syndrome-conditioned logical channels from the identity
\begin{equation}
    P^L:=\sum_sp_s||\fr{p_s}\mathcal{E}_s-I||_\diamond\,,
    \label{e:error_rate}
\end{equation}
and use this as an overall measure of performance of the surface code under those conditions. In practice, we compute this quantity via Monte Carlo sampling of the syndromes $p_s$. It is possible to sample from $p_s$ for non-Pauli noise in the tensor-network algorithm by sequentially contracting the check projectors with the state, then computing the resulting norm of the state by approximately contracting the square-lattice tensor network, as described in Ref. \onlinecite{darmawan_tensor-network_2017}. Note that, due to the use of Monte Carlo sampling, statistical error in $P^L$ from finite sample sizes will be present even if $\mathcal{E}_s$ are computed exactly.

Note that when the number of encoded qubits $k$ is increased from one (by, for instance, modifying the boundaries or adding additional holes to the surface), the logical channel can be described by a $4^k\times 4^k$ matrix, which is exponentially large in $k$. This, however, does not rule out the possibility of efficient decoding using the above method, since it is in principle possible to all but trace out all but one qubit and decode that qubit using a reduced logical channel (represented as a $4\times 4$ matrix)  for that qubit alone. This is analogous to the technique of marginalization described in \cite{darmawan_low-depth_2022} for tensor-network maximum-likelihood decoding. 

\subsection{Tensor-network decoder}
\label{s:tndecoder}
In the previous section we showed that if the logical channel $\mathcal{R}_s\circ \mathcal{N}$ can be calculated, then optimal decoding is possible. In this section we explain how the logical channel can be represented as a tensor-network contraction. This was shown previously in Refs. \cite{darmawan_tensor-network_2017, darmawan_linear-time_2018} which contain more explicit details of the construction. Here we assume syndrome extraction is noiseless , however we provide some details of how circuit level noise can be treated in Sec. \ref{s:optdecoder}. 

The appeal of this decoder is that it is very simple to optimally adapt to a nearly arbitrary noise map, provided that the noise map has an efficient description as a two-dimensional tensor network. As a special case, the decoder can be optimally adapted to arbitrary local noise (in particular non-Pauli noise). Adaptation to correlated noise is also possible, and was demonstrated in Ref. \onlinecite{darmawan_linear-time_2018}, however we do not consider that class of noise here. To summarise the main idea, the decoding algorithm involves constructing a square-lattice tensor network using information from the syndrome and the suspected noise map, then contracting the network using the time evolving block decimation (TEBD) algorithm\cite{vidal_efficient_2003, schollwock_density-matrix_2011}. 

\subsubsection{Tensor network states}
Here we provide a basic description of tensor network states. More information can be found in a variety of pedagogical resources~\cite{orus_practical_2014-2, bridgeman_hand-waving_2017, biamonte_lectures_2020}. A tensor network can be visualised as a graph, where every vertex in the graph corresponds to a tensor, which is a multi-index array of complex numbers. The number of edges incident to a vertex corresponds to the number of tensor indices so, for instance, a matrix would correspond to a vertex with two incident edges. The dimension of an index is called the bond dimension. When two vertices are connected by an edge, the corresponding pair of indices are summed over (contracted), just as pairs of indices from two matrices are summed over in matrix multiplication. A vertex of a tensor network can have an incident edge which is not connected to another vertex and we call such indices physical indices. These indices are not contracted. In contrast, the  edges that are contracted are called virtual indices. A many-body quantum state is called a tensor-network state if the coefficient tensor $\psi_{i_1, i_2, \dots, i_n}$ of the many-body wave function expressed in a local basis $\sum_{i_1, i_2, \dots, i_n}\psi_{i_1, i_2, \dots, i_n}\ket{i_1, i_2, \dots, i_n}$ can be expressed efficiently as a tensor network, where the indices $i_1,\dots,i_n$ are physical indices. 

\subsubsection{Computing the logical channel with tensor networks}
\begin{figure}
    \centering
    \includegraphics[width=0.4\textwidth]{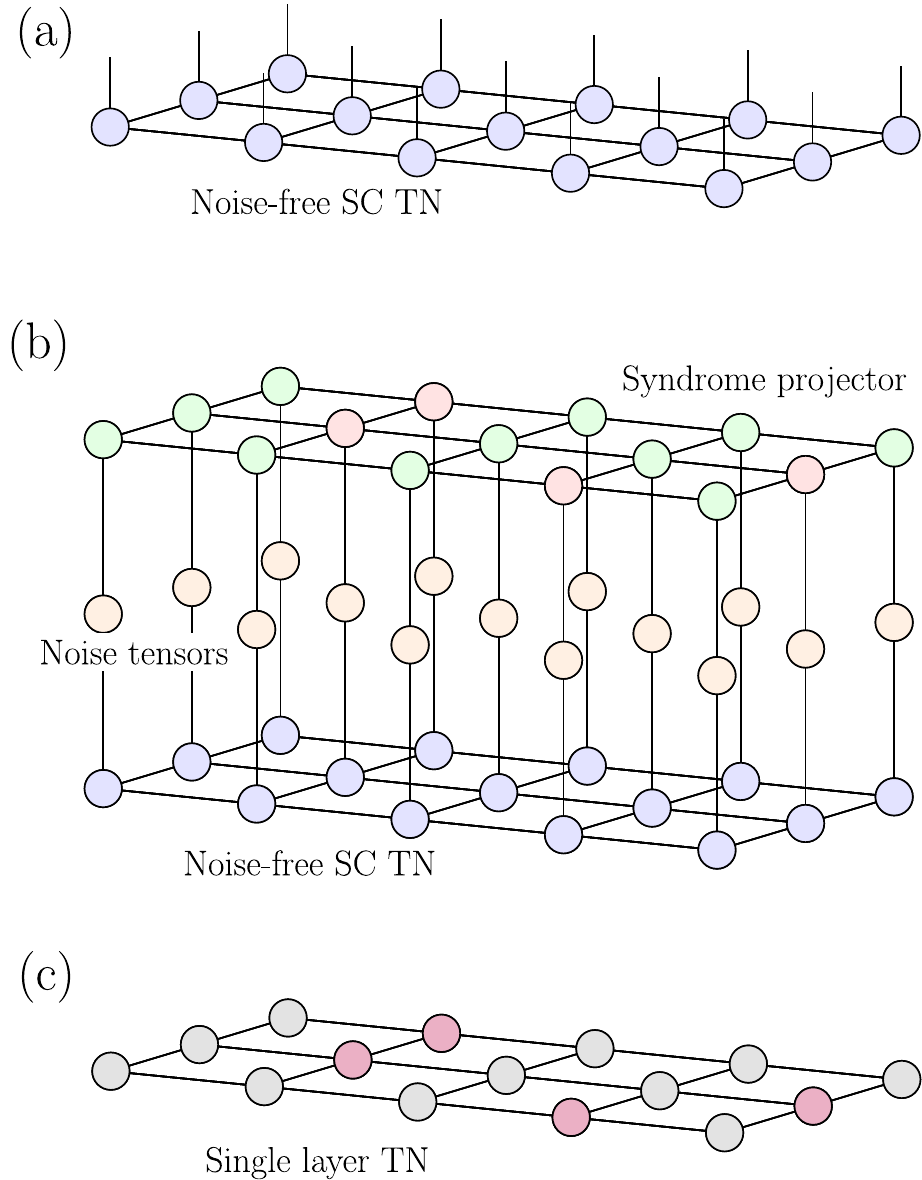}
    \caption{(a) A square-lattice PEPS representing a SC code state, with one physical index per tensor. The density operator for this state can be represented by a TN with the same form, but with physical and virtual bond dimensions squared. (b) A TN representing an entry $C_{ij}$ of the Pauli transfer matrix of the logical channel for the surface code under a local noise map with noiseless syndrome extraction. The network essentially has three layers corresponding to the noiseless surface code state, local noise and a syndrome projector, with the physical indices traced out. (c) The same quantity, expressed as a single layer TN, which is obtained from (b) by contracting the three layers vertically. Specific details on how to define the tensors in these tensor networks are provided in Ref. \onlinecite{darmawan_tensor-network_2017}.}
    \label{f:sc_tn}
\end{figure}

The surface code wave function for any encoded state can be described as a type of tensor-network state called a projected entangled pair state (PEPS) \cite{schuch_peps_2010}. The graph of the PEPS describing the surface code is a square lattice, where each tensor can be thought of as sitting at the location of a physical qubit, as illustrated in Fig. \ref{f:sc_tn}(a). Each tensor has a single physical index and is connected to its nearest neighbours with an edge of bond dimension 2. One particular way of defining the tensors in the TN that is convenient for the purposes here is provided in Ref. \onlinecite{darmawan_tensor-network_2017}. The PEPS structure of a code state $\ket{\psi}$ implies that the density operator $\rho=\ketbra{\psi}{\psi}$ can be expressed simply as a projected entangled pair operator (PEPO), which has the same structure as the PEPS, but with the bond dimensions of virtual indices squared and with two physical indices per tensor. This PEPO can be viewed as a PEPS by merging the two physical indices to a single index with squared bond dimension. 

In the case of noiseless syndrome extraction, the logical channel $\mathcal{R}_s\circ \mathcal{N}$, before applying the decoder correction, can be calculated by computing all 16 elements of the Pauli transfer matrix
\begin{equation}
    C_{ij}={\rm Tr}(L_i (\mathcal{R}_s\circ \mathcal{N}(L_j)))\,,
    \label{e:pauli_transfer}
\end{equation}
where $L_{i/j}\in \{\overline{I}, \overline{X}, \overline{Y}, \overline{Z}\}$ are logical Pauli operators and while the operators $\mathcal{R}_s$ and $\mathcal{N}$ are applied as operators to the density matrix, the $L_i$ is simply left matrix multiplied to it. 

The tensor-network structure of the code states can be exploited to calculate $C_{ij}$. We first start with an PEPS description of a (generally non-physical) code `state' $\rho=L_j$. For $L_i\in \{\overline{X}, \overline{Y}, \overline{Z}\}$ this is not a physical state, but nevertheless has an effective PEPS description (in Ref. \onlinecite{darmawan_tensor-network_2017} and supplementary material, it is explained how PEPS descriptions of arbitrary encoded states can be obtained easily by considering an encoded state entangled in a Bell pair with an unencoded ancilla).  
The application of local noise $\mathcal{N}$, the recovery $\mathcal{R}_s$ and the  logical Pauli $L_i$ to the encoded state all preserve its PEPS structure, and thus $L_i(\mathcal{R}_s\circ \mathcal{N}(L_j))$ can also be expressed as a PEPS with multiple layers. Evaluating the trace of this quantity, thereby computing $C_{ij}$, involves tracing out pairs of physical indices resulting in a tensor network with no physical indices, as illustrated in Fig. \ref{f:sc_tn}(b).  

This tensor network can be compressed into a single layer by contracting along the time dimension, as illustrated in Fig. \ref{f:sc_tn}(c). The resulting square-lattice tensor network can then be contracted using established methods. One method (see Ref. \onlinecite{schollwock_density-matrix_2011} for details) involves treating the left-hand column of the network as a matrix product state (MPS): a tensor network describing a one-dimensional chain. This chain evolves from left to right as columns (matrix product operators) are applied to it. To prevent exponential growth in the bond dimension of this one-dimensional state, an approximation is required. For every applied column, the singular value decomposition is used to truncate the bond dimension of the resulting MPS to a specified value $\chi$, with larger $\chi$ resulting in a more accurate approximation (which becomes exact in the limit as $\chi\rightarrow \infty$). 

Alternatively the square lattice tensor network can be contracted exactly by merging all of the tensors in the left column into a single tensor, with many outgoing indices, then contracting tensors from the remaining TN column by column starting at the top of each column then proceeding down. The memory cost of this method scales exponential with the number of tensors (equivalently, physical qubits) in the left column, which is an improvement over exact state-vector simulations which have a cost exponential in the total number of qubits. 

The above contraction methods will produce $C_{ij}$ representing the logical channel $\mathcal{R}_s\circ \mathcal{N}$ or an approximation of it for a given syndrome $s$. Given this information, one can then evaluate the right hand side of Eq. \eqref{e:optimal} by composing the this channel with each of the four logical Pauli operators, and computing the diamond distance of each resulting operator to the identity (the diamond distance can be computed using convex optimisation \cite{watrous_semidefinite_2009}). The optimal correction, as defined in Eq. \eqref{e:optimal} is the logical Pauli operator resulting in the smallest diamond distance from the identity.

\subsection{Other optimal surface-code decoders}
\label{s:optdecoder}
The above tensor-network decoder can be applied to arbitrary local noise in the case of noiseless syndrome extraction. In certain cases, other implementations of optimal decoding may be used. For example, if the noise is local Pauli noise, the tensor-network decoder of Ref. \onlinecite{bravyi_efficient_2014-1} is also optimal if the tensor network is contracted exactly. The decoder of Ref. \cite{bravyi_efficient_2014-1} is more efficient than the simulation-based decoder used in this work for Pauli noise because the tensor network that needs to be contracted has a lower bond dimension. We use this decoder to study biased noise in Sec. \ref{s:bias}.

The generalization of these tensor-network decoders to the situation where the syndrome readout circuit is noisy is more challenging. Circuit-level decoding by contracting three-dimensional tensor networks has recently been carried out \cite{piveteau_tensor_2023}, where the approach was shown to have high (albeit somewhat sub-optimal) accuracy relative to other decoders. In this work, when taking into account noisy syndrome extraction, we use a simpler, but less computationally efficient optimal decoder that computes the logical channel using full state-vector simulation. In other words, the exponentially large list of density-matrix coefficients is stored in memory without approximation, and it is updated as operations are applied to it.

The logical channel takes a similar form to the noiseless syndrome extraction case, except that noisy syndrome measurements are applied $K\ge1$ times before the final recovery, so the process matrix of Eq. \eqref{e:pauli_transfer} becomes
\begin{equation}
    C_{ij}={\rm Tr}(L_i (\mathcal{R}_{s_{K+1}}\circ \left[\prod_{i=1}^K \Pi_{s_i}\right](L_j)))\,,
\end{equation}
where $\Pi_{s_i}$ is the noisy syndrome projector in the $i$-th round. Note that here, noise is incorporated into the $\Pi_{s_i}$ operators unlike in Eq. \eqref{e:pauli_transfer}, where $\Pi_s$ is separated from the noise $\mathcal{N}$. The final recovery $\mathcal{R}_{s_{K+1}}$ is the same as the noiseless recovery defined in Sec. \ref{s:optdecoder}. Therefore $C_{ij}$ represents the Pauli transfer matrix of a single-qubit channel, as in the noiseless syndrome extraction case. Assuming noiseless recovery can be justified by the fact that a logical readout of the surface code can be performed with single-qubit measurements, which can be regarded as noiseless, since any single-qubit measurement error can be regarded as having occurred before the measurement. Since state-vector simulations are computationally expensive, we have restricted to $d=3$ codes for noisy syndrome extraction simulations, and have also set $K=3$. 

We remark that other decoding algorithms exist that can be optimally adapted to certain types of noise and could in principle be used for the same purpose as the decoders described above. These include provably efficient algorithms for pure X, Z noise \cite{bravyi_efficient_2014-1} and efficient simulation algoriths for simulating coherent X and Z unitary rotation \cite{bravyi_correcting_2018} (which can be used to compute the logical channel for decoding the same noise). Both of these algorithms rely on mapping the problem into a fermionic linear optics simulation. Another tensor-network method for optimal decoding involves computation of the weight enumerator polynomials \cite{cao_quantum_2023}. This technique has been demonstrated to i.i.d local noise (including non-Pauli noise).

\subsection{Identifying critical noise parameters using simulation-based decoding}

We now explain how to use our implementation of an optimal surface-code decoder to determine the critical noise parameters for decoding. Let's say that we have a surface code with a local physical noise map $\mathcal{N}$. The decoder that achieves optimal performance for this noise map $\mathcal{D}_{\rm opt}$ can be implemented using the algorithms described above. Now consider a miscalibrated decoder, i.e. a decoder $\mathcal{D}'$ rather than $\mathcal{D}_{\rm opt}$ that is optimal for a different noise model $\mathcal{N}'\ne\mathcal{N}$. This miscalibrated decoder can be implemented using the same algorithm as the optimal one but using different parameters $\mathcal{N}'$ rather than $\mathcal{N}$ for the decoder noise model. We will in general observe a drop in performance when using $\mathcal{D}'$ rather than $\mathcal{D}_{\rm opt}$ when the physical noise is $\mathcal{N}$, however the magnitude of the drop will depend on how $\mathcal{N}$ and $\mathcal{N}'$ differ. By varying the decoder noise model $\mathcal{N}'$ with the physical noise $\mathcal{N}$ fixed and observing the resulting drop in performance, we can learn which noise parameters decoding is sensitive to. If a single noise parameter in $\mathcal{N}'$ and $\mathcal{N}$ differs by a large amount, yet the logical error rates obtained with $\mathcal{D}'$ rather than $\mathcal{D}_{\rm opt}$ are nearly identical, then the mischaracterized noise parameter likely does not need to be taken into account to achieve optimal decoding. Conversely, if the logical error rates differ significantly, then the noise feature associated with this parameter, must be accurately taken into account by the decoder. 

Thus, by simulating the surface code under a given noise model, and by selectively mischaracterizing various noise features in the decoder noise model, we can identify which which noise features have the biggest impact on decoder performance. In the following sections we describe how this approach can be applied to determine critical noise parameters for a variety of noise types.

\section{Noise-adapted decoding for various noise types}
We have used the above approach to study the potential benefit of noise-adapted decoding for a variety of noise features. 
The noise features that we focus on are coherence, inhomogeneity and bias which we describe in the following sections. A more general approach, which could be applied to an arbitrary single-qubit noise model, is demonstrated in Sec. \ref{s:generic}.

\subsection{Inhomogeneity}
\label{s:inhomogeneity}
Noise in many quantum information processing architectures is spatially inhomogeneous, i.e. it varies across different qubits in the device. Here we explore the extent to which adapting a decoder to the spatial inhomogeneity in the noise can provide a performance improvement. 

We consider a realistic inhomogeneous noise model of amplitude-phase damping. In superconducting architectures, amplitude damping and phase damping are dominant sources of error, and the evolution of the qubit density matrix under both types of noise simultaneously is given by
\begin{align}
    \mathcal{N}_{\rm APD}(\rho)&=\left(\begin{array}{cc}\rho_{00}+\gamma\rho_{11}&\rho_{01}\sqrt{(1-\gamma)(1-\lambda)}\\
        \rho_{10}\sqrt{(1-\gamma)(1-\lambda)}&\rho_{11}(1-\gamma)\\
        \end{array}\right)\,,\\
            &=\left(\begin{array}{cc}1-\rho_{11}e^{-t/T_1}&\rho_{01}e^{-t/T_2}\\
        \rho_{01}^*e^{-t/T_2}&\rho_{11}e^{-t/T_1}\\
            \end{array}\right)\,,\label{e:adpd}
\end{align}
where $\gamma$ and $\lambda$ can be regarded as the damping probability and photon scattering  probabilities respectively. These are related to the relaxation time $T_1$ and the dephazing time $T_2$ by $e^{-t/T_2}=\sqrt{(1-\gamma)(1-\lambda)}$ and $e^{-t/T_1}=1-\gamma$. According to these definitions, $T_1$ and $T_2$ are constrained by $T_2\le 2 T_1$ where equality holds when the noise is purely amplitude damping (i.e. $\lambda=0$). Without loss of generality, we set $t=1$. We point out that, since this is not a Pauli channel, it is not possible to efficiently simulate within the stabilizer formalism. Hence our simulations of this noise are performed using our exact TN method for distances up to $d=9$.

In experiments, $T_1$ and $T_2$ times can vary significantly over different qubits in a device. In our simulations, we assume that $T_1$ and $T_2$ times vary according to a Gaussian distribution with a cutoff at 0 and with $T_2$ bounded by $T_2\le 2T_1$. For every run of the simulation, $T_1$ and $T_2$ are randomly sampled independently for each qubit from a distribution with fixed mean and variance. We fix the relative standard deviation $\sigma/\mu$ of the $T_2$ distribution to be $0.44$ which corresponds to the variance observed on a real device, Rigetti 19Q \cite{otterbach_unsupervised_2017}. 
We set the mean and variance of the $T_1$ distribution to be the same as $T_2$.

\begin{figure*}[tp]
    \includegraphics[width=\textwidth]{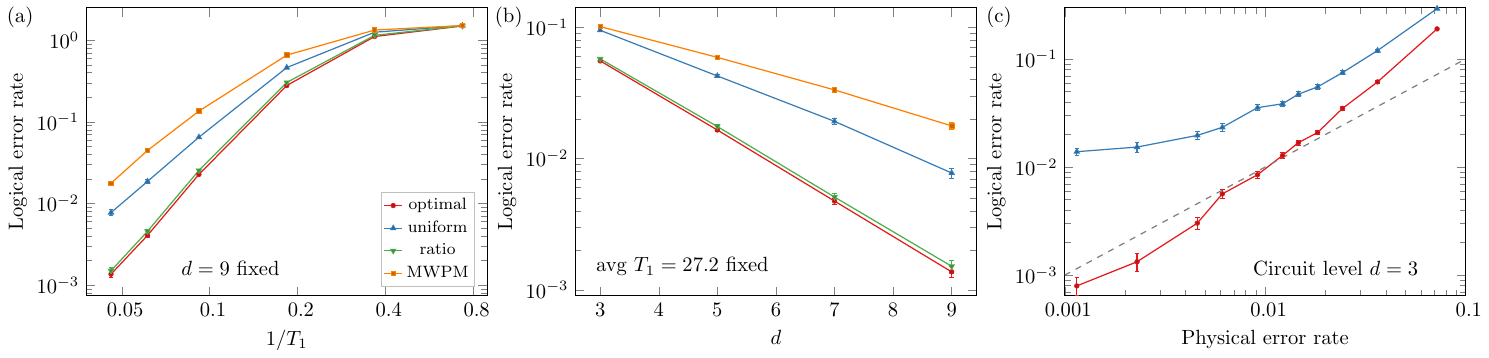}
    \caption{Decoder performance of the surface code with spatially varying amplitude-phase damping noise with $T_1$ and $T_2$ chosen randomly and independently on each qubit from a Gaussian distribution. In (a) and (b) noiseless syndrome extraction is assumed, while in (c) noise is considered on both ancilla qubits as well as data qubits in the syndrome measurement circuit. The logical error rate obtained with an optimal decoder (red) is plotted along with various suboptimal decoders: one which is adapted to a homogeneous noise model with same average $T_1$ and $T_2$ values (blue), one which has accurate information only about the ratio of $T_1$ to $T_2$ on each qubit (green) and standard MWPM with a uniformly weighted syndrome graph (orange). In (a) and (c) the code distance $d$ is fixed (to 9 and 3 respectively) while average $T_1$ is varied. In (b), the code distance is varied while average $T_1=T_2=27.2$ is fixed. In all plots, error rate is quantified by the average diamond distance from the identity. In (c) the dotted line marks equal logical error and physical error rates, so that break-even is achieved below this line. } 
    \label{f:inhomo}
\end{figure*}

The results of our simulations with different decoders are shown in Fig. \ref{f:inhomo}. Four decoders are considered. The first decoder is optimal, meaning that it is perfectly adapted to the $T_1$ and $T_2$ of every qubit. In other words, the decoder noise model is exactly equal to the physical noise model. This corresponds to the case where noise on every qubit is perfectly characterized, and the decoder uses this information optimally. 

The second decoder `uniform' only has knowledge of the average $T_1$ and $T_2$ of each qubit, i.e. the decoder noise model is homogeneous amplitude-phase damping with $T_1$ and $T_2$ on each qubit equal to the distribution averages. In other words, the uniform decoder noise model is a spatially homogeneous approximation to the true noise model.  

The third decoder `ratio' only has accurate information about the ratio of $T_1$ and $T_2$ on each qubit but not the actual values. To be precise, the decoder noise model of this decoder is the same as the physical noise model, except that, for each qubit, $T_1$ and $T_2$ are both multiplied by the same random number from the interval $(0.5, 2)$, which is sampled uniformly and independently for each qubit and each run of the simulation. In other words, `ratio' is adapted to a noise model with the same the ratio of $T_1$ and $T_2$ as the physical noise model on each qubit, however, the actual values of $T_1$ and $T_2$ in the decoder noise model may be off by up to a factor of two. The fourth decoder is standard minimum-weight perfect matching using a uniformly weighted syndrome graph.

The results in Fig. \ref{f:inhomo}(a) and (b) are for noiseless syndrome extraction, in which all four types of decoder are considered. In Fig. \ref{f:inhomo}(c) we show the results of noisy syndrome extraction using a somewhat simplified amplitude-phase damping circuit noise model in which optimal and uniform decoders are considered. In the circuit noise model, each ancilla and data qubit has a randomly sampled $T_1$ and $T_2$ (from the distribution described above) and the amplitude-phase damping channel with corresponding $T_1$ and $T_2$ is applied to each data qubit before every round of syndrome measurements and to ancilla qubits after every initialization. 

In Fig. \ref{f:inhomo}(a) and (b) we observe a large drop in performance for the uniform decoder that is not given information about the noise inhomogeneity. 
The large increase in logical error rate observed with the uniform decoder as a function of code distance (of around an order of magnitude for the largest system size) suggests that significant savings in overhead can be obtained with a decoder that correctly accounts for the inhomogeneity. If we extrapolate the lines in Fig. \ref{f:inhomo}(b), we see that to achieve an error rate of $10^{-10}$, a distance of roughly $d\sim35$ would be required when the decoder is optimally adapted to the noise, compared to $d\sim 52$ for the uniform decoder. This corresponds to a saving of more than a factor of two in the required number of physical qubits, and is a result of simply using more information about the noise in the classical processing.   

A similar trend is also seen for circuit level inhomogeneous noise in Fig. \ref{f:inhomo}(c). In this case, we have expressed the physical noise strength in terms of the diamond norm distance from the identity of the amplitude-phase damping channel with $T_1$ and $T_2$ equal to the distribution average. Then the straight line $y=x$ represents the break even point, so that, roughly speaking, below this line the logical qubit has a lower error rate than the physical qubit. We see that, for this noise model, break even can be achieved at a physical error rate of around $\sim 1\%$ using the optimal decoder, compared to the uniform decoder which does not approach break even when the physical error rate is an order of magnitude lower. 
Of course, in a physical experiment, many more noise processes will be present, however these examples show that adapting decoders to realistic variation on qubits can result in a rather large performance improvement. This results agree with other studies that have shown that adapting variants of MWPM to noise inhomogeneity \cite{iolius_performance_2022-2, iolius_performance_2023, carroll_subsystem_2024} can lead to a performance improvement.

Our results also show that complete information about the $T_1$ and $T_2$ on each qubit is not necessary in order to achieve near-optimal decoding performance. We see in Fig. \ref{f:inhomo}(a) and (b) that the performance of the ratio decoder is extremely close to optimal. The ratio decoder does not have accurate information about the $T_1$ and $T_2$ on each qubit, only the ratios of the two. 
This demonstrates that the decoder can perform optimally with far less than complete information about the noise. The fact that only the ratio between $T_1$ and $T_2$ needs to be known accurately agrees with Ref. \cite{hammar_error-rate-agnostic_2022}, in which high performace was achieved by a decoder that only considered the relative probabilities of phase-flip and bit-flip errors in a spatially homogeneous Pauli noise model.

\subsection{Coherence}
\label{s:coherence}
Theoretical studies of quantum error correcting codes typically assume a Pauli noise model, where errors consist of random Pauli errors drawn from a probability distribution. For stabilizer codes such as the surface code and, more generally, Clifford circuits, this type of noise is convenient since it can be efficiently simulated classically within the stabilizer formalism \cite{gottesman_stabilizer_1997}. 
However, it is rare that this assumption will exactly hold in physical systems. Common types of non-Pauli noise include amplitude damping noise, which we examined in the previous section, and systematic over rotations (caused by imperfect gate control). 
\begin{figure}
    \includegraphics[width=0.52\textwidth]{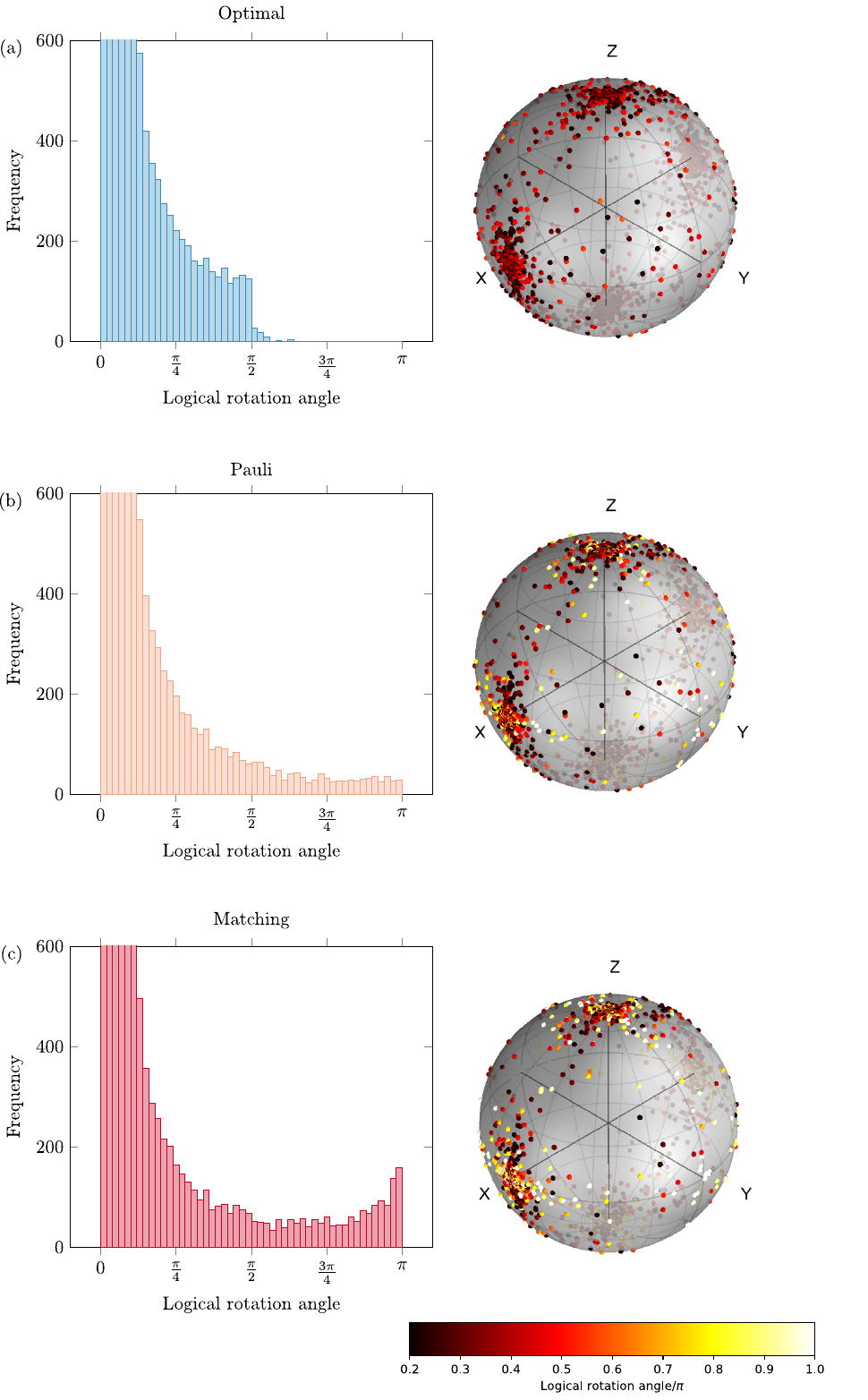}
    \caption{Logical rotation angle $\theta_s$ and axis $\vec{r}_s$ distributions for three different surface-code decoding algorithms where the physical noise on each qubit is a coherent rotation $U(\theta, \vec{r})$ with $\theta=0.2\pi$ and $\vec{r}=1/\sqrt{3}(1,1,1)$ fixed. The surface code distance $d=9$ is also fixed. The sampled axes $\vec{r}_s$ are drawn as points a sphere next to the corresponding angle histogram. On the sphere, each point represents a sampled logical rotation axis $\vec{r}_s$, and the color of each point indicates the rotation angle, with a scale defined by the color bar. Only samples producing rotation angles $\theta_s\ge0.2\pi$ are shown on the sphere. In (a) an optimal decoder is used. In (b) a decoder adapted to the Pauli twirl of the physical noise is used. In (c) standard MWPM using a uniformly weighted syndrome graph is used. The height of the angle-distribution bins close to zero are truncated to on the plot to improve visibility. For reference, the frequency of the bin closest to zero is nearly $2\times 10^4$ for each plot. The same syndrome samples are used for each decoder, so syndromes on which two decoders return the same correction will have identical logical rotation axes and angles.}
    \label{f:bloch_angle}
\end{figure}

Although it is not a universal accepted definition, for simplicity we say noise is coherent if it is not Pauli noise. A simple way to view the coherent component of noise is using the $\chi$ matrix representation of quantum channels. In this representation an $N$-qubit CPTP map $\mathcal{N}$ is expressed as 
\begin{equation}
    \mathcal{N}(\rho) = \sum_{i,j=1}^{4^N} \chi_{ij}P_i \rho P_j 
    \label{e:chi_matrix}
\end{equation}
where $\{P_i\}_{i=1}^{4^N}$ is the set of Pauli operators on $N$ qubits and the positive semi-definite matrix $\chi_{ij}$ is called the $\chi$ matrix. If $\chi_{ij}$ is a diagonal matrix, $\mathcal{N}$ is a Pauli channel and the diagonal element $\chi_{ii}$ represents the probability of the Pauli error $P_i$. For a general channel, we call the off-diagonal elements of $\chi_{ij}$ the coherent terms and diagonal elements the incoherent terms. Previous studies have shown that the performance of the surface code under coherent noise can be worse than under a Pauli twirled version of the noise \cite{darmawan_tensor-network_2017, bravyi_correcting_2018}. In other words, the Pauli twirl approximation can underestimate the effect of the noise. 

Although methods for characterizing coherent noise have been proposed \cite{kaufmann_characterization_2023}, many efficient methods for noise characterization don't measure the coherent part of the noise~\cite{emerson_scalable_2005, magesan_scalable_2011, magesan_efficient_2012, harper_efficient_2020, flammia_efficient_2020, flammia_pauli_2021, harper_fast_2021, flammia_averaged_2021, wagner_learning_2023-1}. So while two noise models may look identical according to incoherent characterization methods, the performance of the surface code under these noise models may be quite different if one of the noise models has coherence.

In this section we will investigate optimal decoding of coherent noise. As mentioned above, it is known that coherent terms can significantly affect surface-code performance, however the effect on decoding specifically has not been previously explored. We consider two noise models, a single-qubit unitary rotation, and a systematic overrotation in the application of a $\rm CNOT$ gate.

\subsubsection{Unitary rotations}
\label{s:unitary_rotations}
First, we explore the decoding of coherent noise using the simple non-Pauli noise model given by a single-qubit unitary rotation, which is defined as 
\begin{equation}
    U(\theta, \vec{r}) = \exp(i\theta\vec{r}\cdot \vec{\sigma}/2 )\,,\label{e:rotation}
\end{equation}
where $\vec{\sigma} = (X, Y, Z)$ is a vector of Pauli matrices, $\vec{r}=(r_x, r_y, r_z)$ is a unit vector representing the axis of rotation and $\theta$ is the angle of rotation.

The case where $\vec{r}=(0,0,1)$, i.e. a rotation about the $z$ axis, was studied in detail in in previous works \cite{bravyi_correcting_2018, venn_error-correction_2020, venn_coherent_2022, behrends_surface_2022, marton_coherent_2023}. Our TN simulation algorithm allows us to study arbitrary rotation axes $\vec{r}$, however approximation in the tensor network contraction are required to make this algorithm efficient in system size. To avoid uncontrolled sources of error, in this study perform exact simulation-based decoding, which limits the size of the surface codes studied to around $d=9$. 

For the surface-code layout considered in this work, there is a simple connection to coherence on the physical level and coherence on the logical level. When the physical noise is a product of local unitary rotations (which may vary over the qubits of the code) the logical channel conditioned on a syndrome after a round of noiseless syndrome extraction and decoding is also a coherent rotation. However the rotation angle $\theta_s$ and axis $\vec{r}_s$ of the logical rotation will in general be randomized depending on the observed syndrome and the decoding algorithm employed. We provide a proof of this property in Appendix \ref{s:coherent_to_coherent_proof}. A special case of this was shown in Ref. \onlinecite{bravyi_correcting_2018}, where the authors showed that physical $z$ rotations are transformed to $z$ rotations on the logical level. Our proof of the more general case essentially mirrors the proof $z$ rotations.  Note that since all decoder corrections differ by a Pauli operator, which is unitary, the resulting logical channel conditioned on a syndrome is also unitary regardless of the decoder used. 

Rotations about the principal $x$, $y$ and $z$ axes are special in that the logical rotation axes will be the same as the physical rotation axes. Since physical rotation exactly about a principal axis is a rather special case, we have performed our simulations using a physical rotation about the axis $\vec{r}=1/\sqrt{3}(1,1,1)$, for which the logical rotation axes are randomized. We also investigated physical noise with various rotation axes deviating from the principal axis, however all of them exhibited qualitatively the same behaviour as the $\vec{r}=1/\sqrt{3}(1,1,1)$ axis, and we do not show the results for other axes here.

In the noiseless syndrome extraction case we compare three decoders. The first decoder is optimal, i.e. it returns the logical Pauli that minimizes the strength of the noise on the logical qubit. The optimal decoder can be thought of as determining the correction that yeilds the logical channel with the smallest logical rotation angle for a given syndrome. The second decoder we considered, `Pauli', is provided information only about the incoherent components of the noise. This corresponds to the information that is typically obtained with randomized benchmarking. Specifically, it is optimized to the Pauli twirl of the physical noise model. For the $\vec{r}=1/\sqrt{3}(1,1,1)$ rotation, the decoder noise model for `Pauli' is depolarising noise. This decoder is therefore equivalent to a maximum-likelihood decoder under depolarising noise with error probability $p=\sin^2(\theta/2)$. The final decoder is standard minimum-weight perfect matching applied separately to the $x$ and $z$ syndrome data.

To illustrate qualitative differences between the decoders, we have plotted the resulting distribution of logical rotation angles and axes in Fig. \ref{f:bloch_angle} in the case where the physical noise on each qubit is $U(\theta, \vec{r})$ with $\theta=0.2\pi$ and $\vec{r}=1/\sqrt{3}(1,1,1)$ . This noise strength here is below threshold, in the sense that increasing system size (from $d=3$ to $d=9$) results in an apparent exponential decrease in logical error rate. The overall logical error rates $P^L$ were found to be 0.17, 0.20 and 0.28 for the optimal, Pauli and MWPM decoders, respectively. We see that, while the rotation axes are indeed randomised, they are mainly clustered around the principal $x$ and $z$ axes. Note that the results are presented differently from the histograms in Ref. \onlinecite{bravyi_correcting_2018} in that we have used the convention that a Pauli operator has $\theta_s=\pi$ (compared to $\pi/2$ in Ref. \onlinecite{bravyi_correcting_2018}) and our rotation angle is always defined to be in the range $\theta_s \in[0,\pi]$, since larger rotation angles in the range $\theta_s \in [\pi,2\pi]$ are equivalent to an angle in $\theta_s \in[0,\pi]$ in the opposite direction (i.e. with a flipped rotation axis). 

The difference in the decoders can be seen in the distribution of rotation angles. As can be seen in Fig. \ref{f:bloch_angle}, almost all logical rotation angles sampled with the optimal decoder are less than $\pi/2$.  This comes from the fact that the logical rotation axes are mostly clustered around the $x$ and $z$ axes, and a rotation about any principal axis of angle greater than $\pi/2$ can be reduced to one less than $\pi/2$ in the opposite direction by applying the right Pauli operator (a $\pi$ rotation), which an optimal decoder is able to determine.  Note, however, that logical rotations about axes other than the principal axes cannot necessarily be reduced in this way by Pauli corrections. Since a small number of sampled logical rotation axes are far from the principal axes, some logical rotation angles greater than $\pi/2$ still occur even with optimal decoding. While $\theta_s>\pi/2$ can occur (albeit rarely) we prove in Appendix \ref{s:anglebound} that the logical rotation angle is upper bounded by $\theta_s\le2\pi/3$ when using optimal decoding, regardless of the physical rotation angle. 

In contrast, with the Pauli decoder, a larger number of logical rotation angles are sampled in the whole interval $[\pi/2, \pi]$, compared with the optimal decoder.  This difference shows that a decoder optimized to the full coherent noise obtains a performance improvement over one optimized only to the incoherent component of the noise. 

The distribution of angles produced by the MWPM decoder shows more support on the $[\pi/2, \pi]$ region than the other decoders, indicating suboptimal performance. A more prominent peak also appears at $\pi$. This peak was also noticed in Ref.~\onlinecite{bravyi_correcting_2018} for the $z$ rotation. Our results illustrate that this peak is due to suboptimal decoding, since it cannot appear for the optimal decoder for which the logical rotation angle is upper bounded by $2\pi/3$.

\begin{figure}
    \includegraphics[width=0.40\textwidth]{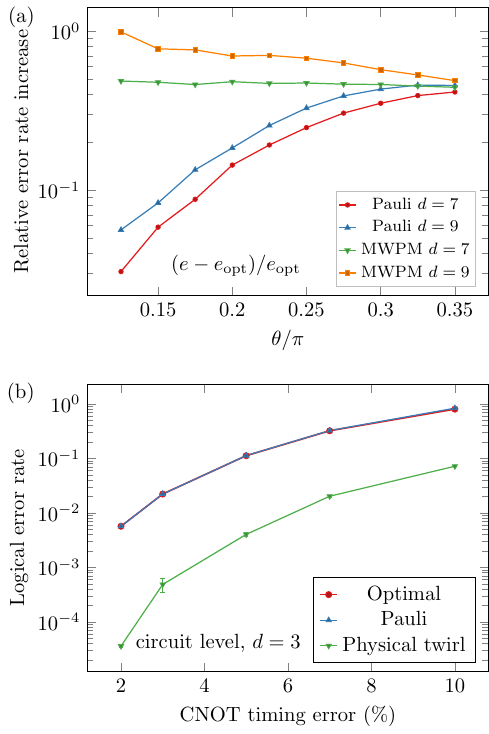}
    \caption{(a) Relative increase in logical error rate due to suboptimal decoding where the physical noise on each qubit is a coherent rotation $U(\theta, \vec{r})$ about the axis $\vec{r}=1/\sqrt{3}(1,1,1)$. The Pauli decoder is adapted to the Pauli twirl of the noise, rather than the full coherent noise map, and the MWPM decoder is standard MWPM using a uniformly weighted syndrome graph. (b) Logical error rate of the surface code versus relative timing error $1-t/t_{\rm CNOT}$ where each $\rm CNOT$ gate is implemented imperfectly due to applying the hamiltonian for time $t$ rather than the ideal $t_{\rm CNOT}$. Here, results are obtained with exact simulation of a $d=3$ surface code in which three rounds of syndrome measurements are performed. The logical error rate obtained using an optimal decoder (red) and a decoder optimized to the Pauli twirl of the noise (blue) are nearly identical when the physical noise is a coherent rotation. The logical error rate obtained by simulating the physically twirled channel (green), on the other hand, is significantly lower.} 
    \label{f:circuit_cohere}
\end{figure}

While the above results demonstrate that adapting the decoder to the full noise map offers some performance improvement over only adapting to the incoherent component, the magnitude of this improvement is small. In Fig. \ref{f:circuit_cohere}(a) we show the relative increase in logical error rate observed using the Pauli decoder and matching decoder compared with the optimal decoder as a function of $\theta$. Here the $y$ axis corresponds to $(e-e_{\rm opt})/e_{\rm opt}$ where $e_{\rm opt}$ is the logical error rate obtained with optimal decoder, as measured by average diamond distance from the identity, and $e$ is the error rate obtained with the indicated decoder (MWPM or Pauli). While the relative increase in error rate of the MWPM decoder increases slightly as $\theta$ decreases, the opposite trend can be seen for the Pauli decoder. As $\theta$ decreases, the performance of the Pauli decoder rapidly becomes close to optimal. For instance at $\theta=0.125\pi$ with $d=9$, the error rates obtained with the optimal and Pauli decoder  $5.8\times 10^{-3}$ vs $6.1 \times 10^{-3}$ respectively: only a $5\%$ relative difference.

\subsubsection{Coherence of the logical channel}
The above suggests that logical channel obtained with the optimal decoder differs qualitatively from suboptimal decoders. Here we explain how with an optimal decoder may tend to increase, rather than decrease coherence in the logical channel (according to a previously used measure of coherence). In Ref.~\onlinecite{bravyi_correcting_2018}, it was shown that for $z$ rotations with MWPM, the angle distribution becomes more sharply peaked at $0$ and $\pi$ as $d$ is increased for a fixed physical rotation angle. For this reason one might say that surface-code error correction decoheres the noise. 

In Ref.~\onlinecite{bravyi_correcting_2018}, a measure of coherence called the coherence ratio was defined as the ratio between the logical error rate $P^L$ (defined in Eq. \eqref{e:error_rate}) divided by the same quantity but with the logical channel replaced with its Pauli twirl, which we call $P^L_{\rm twirl}$. A simple calculation shows that $P^L=2\sum_s p_s |\sin(\theta_s/2)|$ and $P^L_{\rm twirl}=2\sum_s p_s \sin^2(\theta_s/2)$. The coherence ratio $P^L/P^L_{\rm twirl}$ is therefore a measure of coherence, since we have that $P^L\ge P^L_{\rm twirl}$ with equality if and only if $\theta_s$ are all $0$ or $\pi$.

However, a key difference for an optimally noise-adapted decoder is that its angle distribution does not have a peak at $\pi$ (as proved in Appendix \ref{s:anglebound}). 
Furthermore if the noise strength is below threshold, the probability of sampling an angle $\theta_s$ greater than any fixed angle $\theta'>0$ must decrease with increasing $d$ (otherwise, the logical error rate would not be supressed by increasing $d$). This implies that the $P^L_{\rm twirl}$ becomes substantially smaller than $P^L$ and the coherence ratio actually diverges as $d$ is increased. This is the opposite trend to what is observed with MWPM. 

Therefore, for noise that is made up of single-qubit rotations, surface-code error correction with an optimally noise-adapted decoder tends to increase the coherence in the noise, rather than decrease it. We remark, however, that this depends on the definition of coherence. If one considers the syndrome-averaged logical channels, as was also considered in Ref.~\cite{bravyi_correcting_2018} and in other works \cite{beale_quantum_2018, iverson_coherence_2020-1}, rather than the channel conditioned on a syndrome, then different conclusions may be reached.

\subsubsection{Gate-level coherent noise}
Coherent noise is likely to arise in gates due to imperfect control. We have performed small scale density matrix simulations of the $d=3$ surface code to see how coherence affects decoding of gate-level noise. We assume that $\rm CNOT$ gates are implemented by evolving a pair of qubits under the hamiltonian
\begin{equation}
    H = \ketbra{0}{0}\otimes I + \ketbra{1}{1}\otimes X\,,
\end{equation}
for a time $t$ to apply the unitary operator $U=\exp(-itH)$. A perfect $\rm CNOT$ gate is realised if $t=t_{\rm CNOT}:=\pi/2$. Timing errors will result in additional coherent rotation being applied to the qubits. We have simulated the surface code with imperfectly applied $\rm CNOT$ gates, where every gate is implemented with some $t$ less than the ideal value of $\pi/2$. 

In Fig. \ref{f:circuit_cohere}(b) we compare the performance of an optimal decoder to that of a decoder that is optimized to a Pauli twirl of the actual noise map. We find negligible difference in the performance of the two decoders, with only a very small discrepancy occurring at very large noise strengths (at 0.80 for optimal versus 0.83 for the Pauli decoder at when $t/t_{\rm CNOT}=0.9$) and zero discrepancy (within statistical error) observed over the sampled syndromes at the two lowest noise strengths.  

On the same plot we also show the logical error rate obtained when the noise is \emph{physically} twirled. In this case, the noise affecting the qubits is the Pauli twirl of the coherent under-rotation noise defined above, and the decoder is optimized for this Pauli noise model. The error rate is observed for the twirled noise model is more than an order of magnitude lower than for the coherent noise model. 

These results suggest that adapting surface-code decoders to coherent noise terms appears to offer little performance improvement over a decoder optimized to the incoherent noise terms only. Therefore, rather than through coherence-adapted decoding, reducing the coherence of the noise with other methods \cite{hu_mitigating_2022, ouyang_avoiding_2021, debroy_optimizing_2021-1, zhang_hidden_2022, jain_improved_2023-1}, appears to be a more promising way to improve the performance of surface-code error correction with coherent noise. 

\subsection{Bias}
\label{s:bias}
In this section, we use our method to study how information about noise can be used to improve decoder performance when the noise is biased. We say a Pauli noise model is biased if one Pauli error occurs with substantially higher probability than other Pauli errors. Without loss of generality, either $X$, $Y$, or $Z$ can be chosen as the dominant error but, once fixed, the performance of the surface code depends strongly on the local basis in which the checks are defined. We consider the case of single qubit $Y$-biased Pauli noise, where $p_X=p_Z$, $p:=p_X+p_Y+p_Z$, and $\eta:= p_Y/(p_X+p_Z)$ represents the amount of bias towards $Y$ errors. As was shown in Refs. \cite{tuckett_ultrahigh_2018, tuckett_tailoring_2019, tuckett_fault-tolerant_2020}, the threshold of the surface code, with checks defined as in Fig. \ref{f:layout}, when $\eta$ is large is extremely high. Here we investigate the extent to which this high performance is dependent on the decoder having accurate information about the noise.

Note that another surface code variant that we do not consider here, the XZZX surface code, appears to have an even higher threshold when the noise is biased towards $Z$ errors, and there are several potential practical advantages of this variant \cite{bonilla_ataides_xzzx_2021, darmawan_practical_2021-1}. Differences in decoder adaptation for surface code variants and other codes is an interesting topic, however is out of the scope of the current work. 

We simulate the surface code under Pauli noise with different biases using optimal and miscalibrated decoders. As the noise is Pauli, efficient simulation of the surface code is possible using the stabilizer formalism. Furthermore,  we can perform optimal tensor-network decoding without approximation up to system sizes of around $d=25$ using a modified version of the maximum-likelihood decoder of Ref. \onlinecite{bravyi_efficient_2014-1}, which is described in Ref. \onlinecite{tuckett_tailoring_2019}. 

\begin{figure*}[t]
    \includegraphics[width=\textwidth]{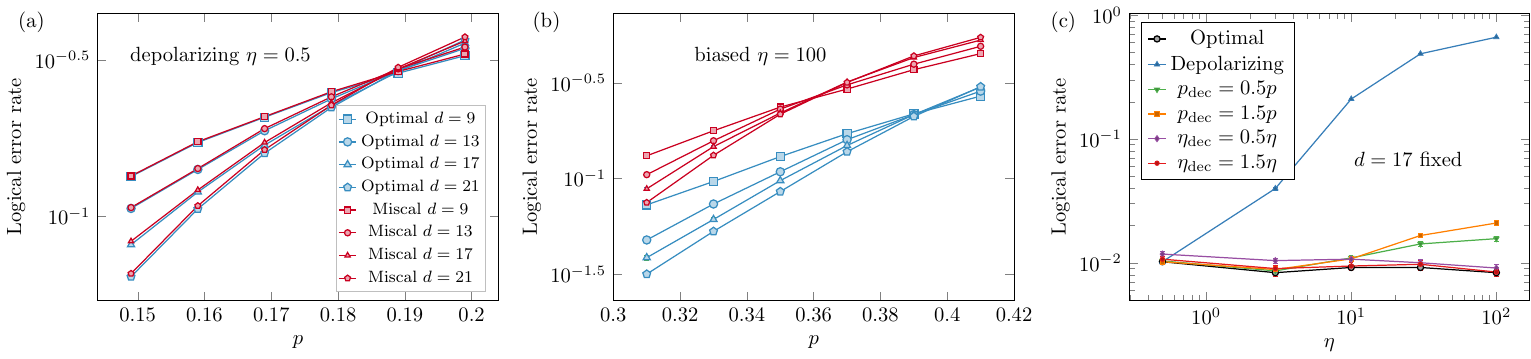}
    \caption{The effect of noise bias on surface-code decoding. In (a) and (b) two decoders are considered: `Optimal' in which the noise model input to the decoder exactly equals the physical noise model, and `Miscal' in which the average infidelity (as typically characterized by randomized benchmarking) in the decoder noise model is mischaracterized $p_{\rm dec} =p/2$.  In (a) the physical noise model is depolarizing noise, while in (b) it is biased noise with $\eta=100$. As can be seen, decoder performance is much more sensitive to miscalibration in the biased noise case. In (c) the physical noise bias $\eta$ is varied, and the optimal decoder is compared against 5 miscalibrated decoders. The physical error rate $p$ is chosen for each value of $\eta$ such that the logical error rate of the optimal decoder is close to $10^{-2}$.   The miscalibrated decoders either use incorrect bias or average infidelity, as indicated. The `Depolarizing' decoder using a depolarizing noise as the decoder noise model (i.e. it assumes no bias). The tensor networks involved in decoding are contracted exactly (effectively $\chi=\infty$).}    
    \label{f:bias}
\end{figure*}

We present the results of these simulations in Fig. \ref{f:bias}. 
In Fig. (a) and (b) we show the results of simulating the surface code under depolarizing (unbiased) noise corresponding to $\eta=0.5$ and biased noise with $\eta=100$ respectively. 

A basic noise property that can be characterized with randomized benchmarking is the average infidelity, which for Pauli noise corresponds to overall error probability $p$. First we consider the case where the average infidelity is underestimated by a factor of two, leading to a miscalibrated decoder. Both a optimal decoder, and a decoder adapted to a noise model that is half as strong, with $p_{\rm dec}=p/2$, are depicted on each plot. 

As can be seen in Fig. \ref{f:bias}(a), for depolarizing noise, little performance is lost when the the average infidelity is severely mischaracterized: The miscalibrated decoder with $p_{\rm dec}$ achieves almost identical logical error rates to the optimal decoder, and the threshold, indicated by the crossing point of the curves of different sizes, barely changes. 

On the other hand, for biased noise with fixed $\eta=100$, there is a noticeable difference in near-threshold error rate between the two decoders. The logical error rate for the largest system size for the miscalibrated decoder is more than a factor of two higher than optimal, and the threshold also decreases slightly from $\sim39\%$ to $\sim 36\%$. 

This suggests that sensitivity to decoder miscalibration increases with the noise bias. In Fig. \ref{f:bias}(c) we present more evidence in support of this. We consider a range of biases from $\eta=0.5$ (depolarizing) to $\eta=100$, and for each bias a noise strength $p$ is chosen such that the logical error rate obtained with the optimal decoder is $\sim 1\%$. For this range of biases we compare several different decoders: an optimal decoder, a decoder optimized for depolarizing noise of the same strength $p_{\rm dec}=p$ (i.e. completely ignorant of the noise bias), and miscalibrated decoders which either  are adapted to a value of $p$ that is off by $50\%$ (i.e. $p_{\rm dec}=1.5p$ or $0.5p$) or a value of bias $\eta$ that is off by $50\%$. 

We see that using no information about the bias in the decoder (the deploarizing decoder) leads to highly suboptimal performance when the bias is large, implying that the performance improvement of the surface code with biased noise is largely tied to having a decoder adapted to the bias. On the other hand, the bias does not need to be known particularly accurately. The decoders with highly miscalibrated bias values ($\eta_{\rm dec}$ off by $50\%$) achieve nearly optimal performance for all values of noise bias considered here. 

Figure \ref{f:bias}(c) also shows that information about error rate $p$ is necessary for optimal decoding at higher biases, in agreement with Fig. \ref{f:bias}(b). A noticeable increase in error rate is observed when $p$ is mischaracterized by 50\%, but only for biases $\eta>30$. Below this, the decoder performs near optimally even for highly miscalibrated $p_{\rm dec}$. 

To summarise, we see that increasing noise bias tends to increase the sensitivity of the logical error rate to miscalibrated decoding. Our results suggest that the decoder needs some information about the bias $\eta$ to perform optimally, however, it seems that only a rough approximation of $\eta$ is sufficient. The noise strength $p$ should be characterized somewhat accurately, but only when the noise bias is large. Fortunately $p$ can be readily characterized simply using well known methods\cite{emerson_scalable_2005}. 

While we did not study alternative surface code variants, like the XZZX code~\cite{bonilla_ataides_xzzx_2021}, in this work, some analysis of the sensitivity of the XZZX code to decoder miscalibration was performed in Ref. \onlinecite{darmawan_practical_2021-1} under a physically motivated circuit-level noise model. An interesting finding was that the XZZX code was substantially less sensitive to decoder miscalibration than the standard surface code considered here. Error correcting codes that do not lose much performance when decoders are miscalibrated may have a practical appeal when the noise varies rapidly and is difficult to characterise.

\subsection{Generic local noise adaptation}
\label{s:generic}
In the previous sections we have looked how surface-code performance can be improved by adapting the decoder to specific noise features. Here we describe a more general approach to determining what types of noise parameters should be taken into account by the decoder in order to achieve optimal performance. The tensor-network method is convenient for this purpose, since it allows us to miscalibrate the decoder noise model in a variety of ways to determine which noise features decoder performance is most sensitive to.

Let us assume that the surface code is affected by a physical noise map $\mathcal{N}$. For any such $\mathcal{N}$ an optimal decoder can be defined as in Eq. \eqref{e:optimal} as the one which minimizes the disturbance to the state on the logical level for any syndrome given the noise map. Recall that a decoder for the noise $\mathcal{N}$ is miscalibrated if the decoder noise model differs from $\mathcal{N}$, i.e. the decoder is adapted to a noise model $\mathcal{N'}$ with $\mathcal{N'}\ne \mathcal{N}$. 

In order to determine the importance of various noise parameters to decoding, we systematically mischaracterize the parameters in the decoder noise model in a number of different ways, and look at the associated drop in performance for each type of miscalibration. Although the exact same method can be applied to any local noise model, we demonstrate it on amplitude-phase damping noise, defined in Eq.~\eqref{e:adpd}. Specifically, we simulate the surface code with $\mathcal{N}_{\rm APD}$ noise, and calculate the logical error rate obtained with 16 miscalibrated decoders which are defined by the decoder noise models $(\mathcal{N}_{\rm APD}+\delta V_i)/A$ where $V_i$, $i=0,\dots,15$ are linearly independent perturbations to the decoder noise model, defined below, $\delta>0$ is the strength of the perturbation and $A$ is a normalisation factor (to obey the trace condition). 

There is no obvious choice for defining the perturbations $V_i$, however, it makes sense to choose perturbations corresponding to physically relevant noise parameters. We specify a convenient set of perturbations $V_i$ as follows. We define the $4\times 4$ matrices
\begin{equation}
    \chi_i:=I + P_i,
    \label{e:perturbation}
\end{equation}
where $\{P_i\}_{i=0}^{15}=\{I\otimes I,I\otimes X,I\otimes Y,\dots,Z\otimes Y,Z\otimes Z\}$ are the Kronecker products of pairs of Pauli matrices. We emphasise that these $P_i$ matrices are simply $4\times 4$ matrices, and do not actually act on a pair of qubits (despite the appearance of the Kronecker/tensor product).  The matrix $\chi_i$ specifies the $\chi$ matrix of the perturbation $V_i$, as in Eq. \eqref{e:chi_matrix}. From Eq. \ref{e:perturbation}, it is clear that each $\chi_{i}$ is a positive semi-definite matrix, which implies that $\fr{A}(\mathcal{N}_{\rm APD}+\delta V_i)$ is a valid physical channel for all $i$. A useful property of this set is that the perturbations can be cleanly split into those that purely affect the incoherent components of the noise (i.e. those with diagonal $\chi$ matrices $i=3,12,15$) and those that affect coherent components. It is also clear that the set $\{\chi_i\}_{i=0}^{15}$ is linearly independent, so any single qubit quantum channel can be expressed as a linear combination of $\chi_i$. 

Note that, in order to save computational time with the large number of decoder noise models (one for each $\chi_i$), we use approximate TN contraction to implement the miscalibtrated TN decoders in this section. We set the bond dimension to $\chi=10$, which appears well converged. 

\begin{figure}[t]
    \includegraphics[width=0.4\textwidth]{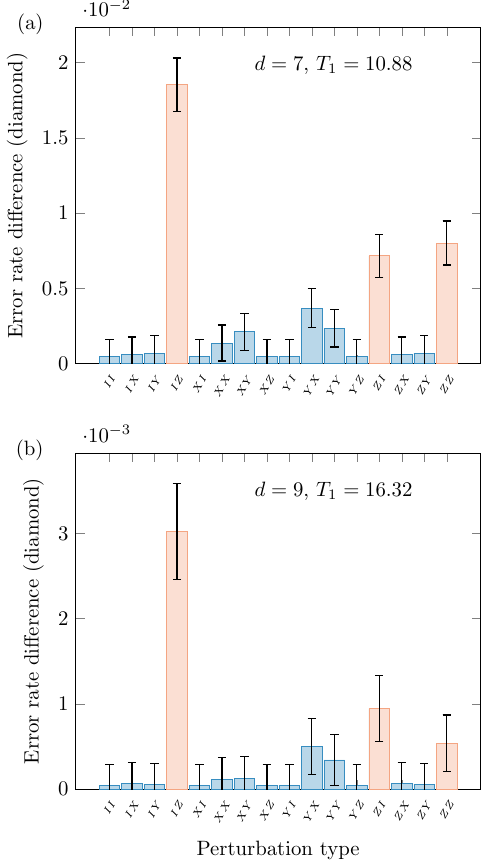}
    \caption{Relative increase in logical error rate (as quantified by the diamond-norm distance from the identity) when the decoder noise model is miscalibrated in 16 different ways. The physical noise model is amplitude-phase damping $\mathcal{N}_{\rm APD}$ with  $T_1=T_2$ and $t=1$. In (a) $d=7$, $T_1=10.88$ while in (b) a larger distance $d=9$ and longer $T_1=16.32$ was chosen, resulting in a lower logical error rate. The logical error of the optimal decoder is $(2.04\pm 0.06) \times 10^{-2}$ in (a) and $(1.0\pm 0.1 )\times 10^{-3}$ in (b). The perturbation type $PQ$ denotes the miscalibrated decoder that uses the decoder noise model $\mathcal{N}_{\rm APD}+0.25V_{PQ}$, where $V_{PQ}$ has a $\chi$ matrix given by $I+P\otimes Q$ as specified in Eq. \eqref{e:perturbation}. Perturbations corresponding to miscalibrating purely incoherent components of the noise (except for a uniform shift) are coloured orange, while the remainder are colored blue. The $IZ$ perturbation, which dominates, corresponds to shifting the probability of $Y$ errors relative to $X$ and $Z$ errors.} 
    \label{f:generic_perturbation}
\end{figure}

In Fig. \ref{f:generic_perturbation} we show the increase in error rate of each miscalibrated decoder compared to the optimal decoder.  The height of each bar is $e-e_{\rm opt}$ where $e_{\rm opt}$ is the logical error rate obtained with the optimal decoder and $e$ is that obtained with the miscalibrated decoder. We present the results for $\mathcal{N}_{\rm APD}$ where both amplitude and phase damping occur with $T_1=T_2$ for different noise strengths and code distances in \ref{f:generic_perturbation}(a) and (b). The case of purely amplitude damping were qualitatively similar and are not shown. 

In both plots, the perturbations with diagonal $\chi_i$ matrices, aside from $\chi_0$ are highlighted orange. They correspond to shifting the incoherent terms of the noise (which can be interpreted as the $X$, $Y$ and $Z$ probabilities) relative to each other. We see that these perturbations have the largest effect on performance. The $I+I\otimes Z$ perturbation, in particular, causes an increases the error rate by approximately a factor of $2$ in (a) and a factor of $4$ in (b). According to the definition Eq. \eqref{e:chi_matrix}, this perturbation corresponds to an increase in the probability of a $Y$ error relative to $X$ and $Z$ errors in the decoder noise model. This is equivalent to perturbing the bias $\eta$ of the noise defined in Sec. \ref{s:bias}. Since a $Y$ error can be thought of as a simultaneous $X$ and $Z$ error, this perturbation can be interpreted as mischaracterizing the correlations between $X$ and $Z$ errors on a single site. Hence, in this case, it appears the noise parameter most important for decoder performance is the $X$ and $Z$ error correlations or, equivalently, the bias $\eta$. 

It can be seen that most of the perturbations corresponding to modifications to the coherent components of the noise (the blue bars) result in little difference in performance. This supports our findings in Sec. \ref{s:coherence}, that for the most part, the coherent component of noise does not need to be accurately characterized in order to obtain near optimal decoder performance.

We emphasise that while we have demonstrated the method on amplitude-phase damping noise, it may be readily applied to other noise models. Furthermore, different sets of perturbations may be chosen depending on what noise features are of interest.

\section{Discussion and Conclusions}
In this paper we present an approach to identifying the critical noise parameters for surface-code decoding. Our approach allows us to target a specific noise parameter and to quantify the cost (in terms of logical error rate) of not adapting a decoder accurately to that parameter. 
We have applied this approach to a number of noise features of practical interest, including spatial inhomogeneity, coherence, and bias. We have also demonstrated a general approach to determining key noise parameters in local noise, without prior knowledge of what parameters may be important. 

In all of the above cases we find that only a small number of noise parameters need be known accurately in order to achieve near-optimal decoder performance. 
This means that, in order to calibrate a decoder, 
noise characterization methods likely only need to focus on certain critical noise parameters. Finding these critical parameters could potentially help guide the development of new decoding algorithms, that are tailored specifically to those noise parameters.

There are a number of directions for future research. While we have focussed on features of uncorrelated noise in this work, correlations are likely to have a large impact on decoder performance. Correlated noise, which is not describable as tensor product of single-qubit noise maps, has many more parameters than uncorrelated noise, and therefore determining which types of correlation most affect decoder performance is likely crucial to optimizing the performance of the surface code~\cite{harper_learning_2023}. Decoders for correlated noise based on tensor networks could be useful for this purpose~\cite{darmawan_linear-time_2018, chubb_statistical_2021}. 

Although we have emphasized the connection to characterization methods, there could be implications of this approach to decoding strategies that do not directly use information about the noise at all. For instance, machine learning inspired decoders often train internal parameters (e.g. weights in neural networks) using large amounts of syndrome data \cite{torlai_neural_2017, krastanov_deep_2017-1, maskara_advantages_2019, varsamopoulos_decoding_2020, ni_neural_2020, davaasuren_general_2020, theveniaut_neat_2021, gicev_scalable_2023, cao_qecgpt_2023, wang_transformer-qec_2023}. The training process can be costly, in part due to the fact that a large number of parameters need to be optimized. It may be possible to substantially decrease the resources required for this type of optimization by using models with fewer parameters. We would expect that a model in which only the parameters corresponding to the critical noise parameters need to be optimized could be trained much more efficiently than a general purpose neural network. 

\begin{acknowledgements}
This work was supported by, or in part by, the U.S. Army Research Laboratory and the U.S. Army Research Office under Raytheon BBN Technologies Subcontract  LBN9512246 under Contract W911NF-14-C-0048. ASD was supported by JST, PRESTO Grant Number JPMJPR1917, Japan and Grant-in Aid for Transformative Research Areas (A) 21H05183. ASD would like to thank Ben Brown and Stephen Bartlett for stimulating discussions during early stages of this project.

A large part of this work was carried out several years ago in collaboration with David Poulin, who tragically passed away before its completion. This paper is dedicated to his memory.
\end{acknowledgements}

\appendix
\section{Proof that logical channel is a unitary rotation when the physical noise is a product of single-qubit unitary rotations}
\label{s:coherent_to_coherent_proof}
Here we show that when the physical noise is a product of single-qubit unitary rotations, the logical channel obtained after a round of noiseless syndrome measurements, is also a unitary rotation. In general, the rotation angle and rotation axis on the logical qubit will depend on the observed syndrome. In Ref. \onlinecite{bravyi_correcting_2018} a proof was provided for the special case where the noise on each qubit is a unitary rotation about the $z$ axis. In this case, the logical channel is also a rotation about the $z$ axis. The proof of the more general case presented here is very similar to that of the special case provided in Ref. \onlinecite{bravyi_correcting_2018}. 

A general single qubit unitary, as in Eq. \eqref{e:rotation}, can be written as

\begin{equation}
    U(\theta, \vec{r}) = \cos(\theta/2)I+i\sin(\theta/2)(\vec{r}\cdot \vec{\sigma})\,,
    \label{e:superposition}
\end{equation}
i.e. a superposition of an identity term with a real coefficient (the first term) and a Pauli term (the second term) with an imaginary coefficient.
A product of single-qubit unitary rotations $\prod_i U_i$ (where $U_i$ are not neccesarily identical), can be expanded out in the Pauli basis as
\begin{equation}
\prod_i U_i=\sum_j \alpha_j P_j\,,
\end{equation}
where in the second sum $P_j$ runs over all $n$ qubit Pauli operators. Given the form of Eq. \ref{e:superposition} each coefficient $\alpha_j$ is either real or imaginary. A syndrome measurement producing syndrome $s$ will collapse the superposition, eliminating Pauli terms that are inconsistent with the syndrome
\begin{equation}
    \Pi_s \mathcal{N}(\rho) \Pi_s = \Pi_s \left(\sum_{i\sim s} \alpha_i P_j\right) \rho \left( \sum_{j\sim s} \alpha_j^* P_j \right) \Pi_s\,,
\end{equation}
where $\sum_{i\sim s}$ implies summation over all indices $i$ such that the error $P_i$ can give rise to the syndrome $s$.
After applying a correction $f_s$ to return the state to the logical space, each $P_i$ will become equivalent to one of four logical operators. Explicitly, in the logical space
\begin{equation}
    f_s\sum_{i\sim s} \alpha_i P_i = c_I \overline{I} + c_X \overline{X}+ c_Y \overline{Y} + c_Z \overline{Z}\,,
\end{equation}
where 
\begin{equation}
    c_L:=\sum_{i\sim s:i\in L} \alpha_i
\end{equation}
where $\sum_{i\sim s:i\in L}$ implies summation over all $i$ consistent with the syndrome $s$ such that $f_s P_i$ is equivalent to logical Pauli $L$ within the code space. Each summation is over all elements of the stabilizer group. Given that each element of the stabilizer group has even weight on the specified surface-code layout, and every non-identity single Pauli term has a purely imaginary coefficient, as in Eq. \eqref{e:superposition}, this implies that all terms in each summand are either purely real or imaginary. Similarly, $\overline{X}$, $\overline{Y}$ and $\overline{Z}$ logical operators all have odd weight, so that if the coefficients of $\overline{I}$ are all real, the coefficients of $\overline{X}$, $\overline{Y}$ and $\overline{Z}$ must all be purely imaginary and vice versa. Hence the logical channel is a rotation
\begin{equation}
    f_s\sum_{i\sim s} \alpha_i P_i = \cos(\theta'/2)I_L+i\sin(\theta'/2)(\vec{r}'\cdot \vec{\sigma})
\end{equation}
where $\vec{\sigma}=(\overline{X}, \overline{Y}, \overline{Z})$ and, if the coefficient of $I_L$ is real, $\theta'/2:=\arccos(c_I)$ and $\vec{r}':=-i/\sin(\theta'/2)\left(c_X, c_Y, c_Z\right)$. If the coefficient is $I_L$ imaginary, simply divide all coeficients by a (physically irellevant) overal phase of $i$ then use the same definitions. 

\section{Maximum logical rotation angle with optimal decoding of unitary noise}
\label{s:anglebound}
Here we show that when the physical noise is composed of local unitary rotations and optimal decoding is used, the maximum rotation angle of the logical channel (which is also a unitary rotation) is $2\pi/3$. As shown in Appendix \ref{s:coherent_to_coherent_proof}, the logical channel, after applying the recovery and before applying the decoder correction, has the form
\begin{equation}
    U(\theta, \vec{r}) = \cos(\theta/2)I+i\sin(\theta/2)(\vec{r}\cdot \vec{\sigma})\,,
\end{equation}
Given that the decoder corrections are restricted to Pauli operators, an upper bound on the logical rotation angle can be found by choosing the worst possible $\theta$, $\vec{r}$ for a Pauli decoder, i.e. a logical rotation that cannot be corrected well by Pauli corrections. More precisely we want to find 
\begin{equation}
\underset{\theta, \vec{r}}\max \underset{L\in \{\overline{I}, \overline{X}, \overline{Y}, \overline{Z}\}}\min || L\circ U(\theta, \vec{r}) -I||_\diamond\,. 
\end{equation}
It is easy to see that this occurs when the coefficients of $I$, $X$, $Y$, $Z$ are all equal in $U(\theta, \vec{r})$, in which case the logical rotation angle is independent of which Pauli correction is applied. This holds when $|r_x|=|r_y|=|r_z|=1/\sqrt{3}$ and $\tan(\theta/2)=\sqrt{3}$, which gives an upper bound on the logical rotation angle of $\theta=2\pi/3$.

%

\bibliographystyle{apsrev4-2}

\end{document}